\documentclass[12pt,a4paper]{article}
\usepackage{jheppub}
\usepackage{dsfont}
\usepackage{amssymb,amsmath}
\usepackage{epsfig}
\usepackage{epstopdf}
\usepackage{latexsym}
\usepackage{graphicx}
\usepackage{subfigure}
\usepackage{booktabs}
\usepackage{bbm}
\pdfoutput=1
\makeatletter
\def\@fpheader{\relax}
\makeatother

\def\cM{{\cal M}}

\def\S{{\cal S}}
\def\A{{\cal A}}

\def\L{{\cal L}}
\def\O{{\cal O}}

\def\S{{\cal S}}

\title{Flavours and Infra-red Instability in Holography}
\author{Arnab Kundu$^{a,b}$}
\affiliation{$^a$Theory Division, Saha Institute of Nuclear Physics, 1/AF Bidhannagar, Kolkata 700064, India.}
\affiliation{$^b$Homi Bhaba National Institute, Training School Campus, Anushakti Nagar, Mumbai 400085, India.}
\emailAdd{arnab.kundu [at] saha.ac.in}

\abstract{Within a phenomenological holographic model in $(4+1)$-bulk dimensions, defined by Einstein-gravity with a negative cosmological constant, coupled to a Dirac-Born-Infeld and a Chern-Simons term, we explore the fate of BF-bound violation for a probe scalar field and a fluctuation mode of the corresponding geometry. We assume this simple model to capture the dynamics of a strongly coupled SU$(N_c)$ gauge theory with $N_f$ fundamental matter, which in the limit $\O \left( N_c \right) \sim \O\left(N_f\right)$ and with a non-vanishing matter density, is holographically described by an AdS$_2$-geometry in the IR. We demonstrate that, superconductor/superfluid instabilities are facilitated and spontaneous breaking of translational invariance is inhibited with increasing values of $\left(N_f / N_c \right) $. This is similar, in spirit, with known results in large $N_c$ Quantum Chromodynamics with $N_f$ quarks and a non-vanishing density, in which the chiral density wave phase becomes suppressed and superconducting instabilities become favoured as the number of quarks is increased.}

\begin{document}

\maketitle
\flushbottom

\section{Introduction}

Strongly coupled phases of Quantum Chromodynamics (QCD) are varied and interesting. Among these, the long-suspected and argued phase of colour-superconductivity\cite{Alford:1997zt} is particularly intriguing in its novelty and its potential detection in the core of neutron stars. However, the strong coupling dynamics of QCD makes it a theoretical challenge to describe and understand this particular phase from first principles.

One may instead adopt large $N_c$ techniques, specially AdS/CFT correspondence\cite{Maldacena:1997re} to make headway with the challenges of strong coupling, at the cost of studying precisely QCD, and subsequently explore a broader issue of colour-flavour locking in large $N_c$ gauge theories. In the large $N_c$ limit, ref.~\cite{Deryagin:1992rw} has argued, the preferred phase is a particular spatially-modulated phase, namely the chiral density wave, instead of a superconducting one. This analysis, however, considers only adjoint degrees of freedom and forgoes the possibility of a matter sector, in general. Specifically, with matter in the fundamental representation, ref.~\cite{Shuster:1999tn} showed that the chiral density wave inducing instability is sub-dominant over the superconducting instability, provided $N_f / N_c > 10^{-3}$, where $N_f$ is the number of fundamental matter. Taking a broad cue from this, it will be interesting to find a simple model where superconducting instability is favoured over a spontaneous breaking of translational invariance.

Within the framework of AdS/CFT, particularly with adjoint matter fields, spatially modulated phases, as a result of spontaneously breaking of translational invariance, have appeared in many avatars, perhaps beginning with \cite{Nakamura:2009tf} in the vector fluctuations, and followed by {\it e.g.}~\cite{Donos:2011bh, Donos:2011qt} and far too many works to enlist here. In most of the models considered so far, which at least consists of Einstein-gravity with a negative cosmological constant and a Maxwell field, the spatially modulated instability is induced by either a non-vanishing topological term in the gravity action, or by additional ({\it e.g.}~scalar) fields. These models can be obtained by consistent truncations of supergravity, or may simply be effective phenomenological models of gravity with various matter sector.

In this article, we consider a particular phenomenological model, in which we introduce a space-filling Dirac-Born-Infeld term as the matter sector. This, by analogy and inspired from explicit D-brane constructions\cite{Karch:2002sh}, can be viewed as introducing fundamental matter in the system\footnote{This simple model may arise from a UV-complete supergravity or stringy construction, such as considering a D$9$-brane in type IIB supergravity. However, we have not checked whether this is precisely true. Moreover, our motivation, for this article, is not to pin down the particular UV-completion, rather whether we may make a clean and sharp statement in a simple enough model. Also, we will primarily explore various fluctuation sectors, which for a D$9$-brane in $10$-dimensions cannot exist.}, with provision of introducing non-vanishing matter density. The specific model we consider is a simple generalization of \cite{Nakamura:2009tf}, but the interpretation in terms of the putative dual quantum field theory comes naturally equipped with the additional ingredients: fundamental matter sector and non-vanishing density. We will demonstrate, with this simple model, that the introduction of non-vanishing $N_f$, along with a density catalyzes an instability towards the formation of superconducting condensate, and inhibits a spontaneous breaking of translational symmetry in the vector fluctuations, in a precise sense that we explore in the article.

The pure gravity action, in this case, represents a conformal field theory (CFT) with purely adjoint matter. Addition of a DBI-term can be interpreted as an exactly marginal deformation of the CFT which changes the central charge.\footnote{This is rather unusual. Typically, central charge of a CFT is independent of marginal deformations. One known exception is for CFTs with boundaries, where the boundary central charge does not coincide with the bulk central charge, and can respond to a marginal deformation\cite{Herzog:2017xha}. In the current case, this happens because of a trivial shift in the cosmological constant and is a consequence of our construction.} Turning on a non-vanishing density corresponds to a relevant deformation and the system flows to an IR AdS$_2$, which is dual to a $(0+1)$-dimensional conformal quantum mechanics. The IR is non-perturbative in $\left( N_f / N_c \right)$\cite{Kundu:2016oxg}, where flux and the radial scale in the geometry are interlocked. Observed from the UV CFT, as defined by our action, the density perturbation drives the flow to the same IR fixed point with a vanishing central charge. Since the IR is extremal, it corresponds to vanishing temperature in the dual field theory and therefore corresponds to one point in the density-temperature parameter space. The apparent adjustable density-parameter in the UV sets the scale where the UV geometry merges into the IR-geometry, and carries no other non-trivial information.

In the absence of a flux, the marginally deformed CFT appears to be more vulnerable to a BF-bound violation for a neutral probe scalar, assuming that it is sensible to scan through a class of theories as a function of $\left( N_f / N_c \right)$. In the extremal AdS$_2$ region, the possibility of BF-bound violation is more prominent, and is also present for a complex charged probe scalar field. The latter has been associated with a spontaneous breaking of a global U$(1)$-symmetry and hence related to the phenomenon of superconductivity (superfluidity, to be precise). On the other hand, the spontaneous breaking of translational invariance that can be induced by a Chern-Simons term in five bulk dimensions, can be obstructed by increasing $\left( N_f / N_c \right)$.

This article is divided into the following sections: In section 2 we introduce the action and the solutions that we subsequently study. We also discuss the role of the flavour back-reaction, measured by $\left( N_f / N_c \right)$ on the corresponding BF-bound for both neutral and charged scalars. In the next section, we discuss in details the fate of spatially modulated instabilities as a function of the back-reaction strength. In section $4$, we conclude. Many details of the fluctuation calculations, including a study of fluctuations around a constant electric field configuration arising from a DBI in Minkowski, are relegated to two appendices: Appendix A and Appendix B.

\section{The Action, the Solutions \& the BF Bounds}

We begin with the following action:
\begin{eqnarray}
&& S_{\rm full} = S_{\rm gravity} + S_{\rm DBI} + S_{\rm topological} \ , \label{actm1} \\
&& S_{\rm gravity} = \frac{1}{2 \kappa^2} \int d^{d+1} x \sqrt{- {\rm det} g} \  \left(R - 2 \Lambda \right) \ , \label{act2} \\
&& S_{\rm DBI} = - \tau \int d^{d+1} x \  \sqrt{- {\rm det}  \left( g + F \right) } \ , \label{act3} \\
&& S_{\rm topological} = S_{\rm CS} =   \frac{\alpha\tau }{3!} \int d^{5} x \ A \wedge F \wedge F  \ , \quad {\rm in} \quad d=4 \ . \label{actm4}
\end{eqnarray}
Here $S_{\rm gravity}$ represents usual Einstein-gravity that is typically dual to the adjoint sector of an SU$(N_c)$-type gauge theory, $S_{\rm DBI}$ corresponds to the action of a brane that represents addition of a fundamental sector in the theory, and $S_{\rm topological}$ represents a generic topological sector, which in $d=4$, is given by {\it e.g.}~a Chern-Simons term, generated by the U$(1)$ gauge field, denoted by $F = dA$, living on the worldvolume of the putative brane. Furthermore, $\kappa$ represents the Newton's constant, $\tau$ represents the ``brane tension", and $\alpha$ represents the Chern-Simons coupling constant. In the limit of small fields, {\it i.e.}~when $F$ is small compared to $g$, $S_{\rm DBI}$ reduces to a simple Maxwell term, $S_{\rm Maxwell} \sim - \frac{1}{4} F_{\mu\nu} F^{\mu\nu}$.

Before proceeding further, we offer a few comments on our choice of units. We work with the following (length) dimension assignments:
\begin{eqnarray}
&& \left[ g_{\mu\nu} \right] = 0 = \left[ F_{\mu\nu} \right] = \left[ \Lambda \right]  = \left[ \alpha \right] \ , \quad \left[ \kappa \right] = \frac{d+1}{2} = - \frac{\left[ \tau \right]}{2} \ , \\
&& \left[ {\rm coordinates} \right] = 1 \ . 
\end{eqnarray}
With the above choice, Lagrangian densities are dimensionless, and henceforth any first integral of motion of the form $\left( \partial \L / \partial F_{\mu\nu} \right)$ is also dimensionless.\footnote{Note that, on a D-brane, the flux comes naturally multiplied by a factor of $(2\pi \alpha')$, where $\alpha'$ is the string tension. Thus, we are implicitly setting $2 \pi \alpha' =1$ above.} Furthermore, we have assigned the cosmological constant to be dimensionless, therefore any bulk dimensionful quantity is subsequently measured in units of the curvature scale of the empty-AdS solution.

The equations of motion are:
\begin{eqnarray}
&& R_{\mu\nu} - \frac{1}{2} \left( R - 2 \Lambda \right) g_{\mu\nu} = T_{\mu\nu} \ ,  \label{ein} \\
&& \partial_\mu \left( \sqrt {-{\rm det} \left( g + F \right)} \ \A^{\mu \nu} \right) + \frac{\alpha}{2} \epsilon^{\nu\alpha\beta\rho\sigma} F_{\alpha\beta} F_{\rho\sigma} = 0 \ , \label{max}
\end{eqnarray}
where 
\begin{eqnarray}
&& \A^{\mu \nu} = - \left( \frac{1}{g+F} \cdot F \cdot \frac{1}{g- F} \right)^{\mu\nu} \ , \label{anti} \\
&& T^{\mu\nu} = \frac{\kappa^2 \tau}{\sqrt{- {\rm det} g}} \left( \frac{\delta S_{\rm DBI}}{\delta g_{\mu\nu}} + \frac{\delta S_{\rm DBI}}{\delta g_{\nu\mu}} \right) = - \left( \kappa^2\tau \right) \frac{\sqrt{- {\rm det} \left( g+F \right)}}{\sqrt{ - {\rm det} g}} \S^{\mu\nu} \ , \\
&& \S^{\mu\nu} = \left( \frac{1}{g+F} \cdot g \cdot \frac{1}{g- F} \right)^{\mu\nu} \label{sym} \ .
\end{eqnarray}
In calculating the above, $\A^{\mu\nu}$ or $\S^{\mu\nu}$ are calculated by treating $g$ and $F$ as matrices, and then using the formulae in (\ref{anti}), (\ref{sym}). All indices above are raised and lowered by the metric $g$.

In general $(d+1)$-bulk dimensions, the general solution is well-known in analytical form. Here we review the general solution. Towards that, first, suppose there is no flux, and therefore the subsequent AdS$_{d+1}$-BH solution can be written with a trivial shift in the cosmological constant:
\begin{eqnarray}
&& ds^2 = L_2^2 \left( - r^2 f(r) dt^2 + \frac{1}{r^2 f(r)} dr^2 + r^2 d\vec{x}_{d-1}^2 \right) \ , \quad f(r) = 1 - \frac{r_{\rm H}^d}{r^d} \ , \label{Schd1} \\
&& L_2^2 = L^2 \frac{d (d-1)}{d(d-1) - 2 \epsilon L^2} \ , \quad \epsilon = \kappa^2 \tau \ , \quad {\rm with} \quad \Lambda = - \frac{d (d-1)}{2 L^2} \ . \label{Schd2}
\end{eqnarray}
Here, a non-vanishing $\epsilon$ introduces a marginal coupling in the putative boundary dual theory, which, from the bulk gravitational point of view, appears to have merely shifted the cosmological constant to: $\Lambda_{\rm eff} = \Lambda +  \epsilon$.\footnote{Note that, {\it a priori} there is no constraint on $\epsilon$, and thus it naively seems that one can change the sign of the ``effective" cosmological constant. However, we believe this is an artefact of out inherently {\it bottom-up} description, and a UV-complete stringy embedding of the model will come equipped with other fields that prevent a change in the signature of the cosmological constant. Thus, we will preserve the sign and subsequently impose an upper limit on $\epsilon < d(d-1) / 2 L^2$. We thank J.~F.~Pedraza, A.~Rovai and J.~Sonner for very useful conversations related to this issue.}

One can analyze a Klein-Gordon field, in the background of (\ref{Schd1})-(\ref{Schd2}), setting $r_{\rm H} =0$. The corresponding BF-bound violation will occur provided $m^2 L_2^2 < - d^2 /4$. Setting $\epsilon =0$, the BF bound reads $m_0^2 \ge - \frac{d^2}{4}$ (in units of $L=1$). The analogous bound, when $\epsilon \not = 0$, yields
\begin{eqnarray}
m_\epsilon^2 L_2^2 \ge - \frac{d^2}{4} \quad \implies \quad m_\epsilon^2 \frac{d(d-1)}{d(d-1) - 2\epsilon} \ge - \frac{d^2}{4} \ . \label{BFDBI}
\end{eqnarray}
In writing the above, we have included a subscript to the mass, to distinguish the cases with vanishing $\epsilon$ with that of the non-vanishing one. A few comments are in order: First, note that, in the $d\to \infty$ limit $m_\epsilon^2 \to m_0^2$, provided $\epsilon / d^2 \to 0 $. Let us now take the lowest stable mass in AdS with $\epsilon=0$, and the LHS of (\ref{BFDBI}) yields:
\begin{eqnarray}
- \frac{d^2}{4} \left( \frac{d(d-1)}{d(d-1) - 2\epsilon} \right) < - \frac{d^2}{4} \ ,
\end{eqnarray}
which implies that in the AdS$_\epsilon$, the lowest stable mode in AdS$_{\epsilon=0}$, is already below the BF bound. Hence, introducing the DBI-matter, which only shifts the cosmological constant,  favours instability. If we view the AdS$_{\epsilon=0}$ background as the UV CFT, the $\epsilon$ deformation takes us to a CFT, which is more vulnerable to, {\it via} a BF-bound violation, instability.

There is another way of reaching the same conclusion. Let us consider the case $m_0^2 = m_\epsilon^2$, in units of the AdS$_{\epsilon=0}$ curvature. Subsequently we compare the AdS$_{\epsilon=0}$ curvature with that of the AdS$_\epsilon$:
\begin{eqnarray}
L_2^2 - 1 = \frac{2\epsilon}{d(d-1) - 2 \epsilon} > 0 \ ,
\end{eqnarray}
which implies that, in the regime $m_0^2 = m_\epsilon^2 > 0$, a non-vanishing $\epsilon$ enhances positivity in mass, while, in the $m_0^2 = m_\epsilon^2 < 0$ regime, a non-vanishing $\epsilon$ facilitates a BF-bound violation. Thus, in brief, from the UV-perspective, introducing fundamental matter, specially when this corresponds to a marginal deformation of the unflavoured CFT, makes it easier to violate the BF-bound.

A similar question can be asked in the infra-red, specially around the flavour-back-reacted conformal fixed points, which are essentially extremal geometries in the bulk. Towards that, we begin by reviewing the general solution\cite{Pal:2012zn, Tarrio:2013tta}. In the presence of flux, the charged black hole solution is given by
\begin{eqnarray}
&& ds^2 = L_2^2 \left( - r^2 f(r) dt^2 + \frac{1}{r^2 f(r)} dr^2 + r^2 d\vec{x}_{d-1}^2 \right) \ , \\
&& f(r) = \frac{L_2^2}{L^2} + \frac{C_2}{r^{d}} - \frac{C_3}{r^{d-1}} \, _2F_1 \left(- \frac{1}{2} , \frac{1}{2d -2 } ; \frac{2d -1}{2d -2} ; - \frac{L_2^{2(d-1)}}{\rho^2} r^{2 (d-1)}\right) \ , \\
&& C_3 = \frac{\alpha_d}{L_2^{d-3}} \left( \rho \epsilon \right) \ , \\
&& A_t(r) = - \frac{1}{d-2} \frac{\rho}{L_2^{d-3}} \frac{1}{r^{d-2}} \, _2F_1 \left( \frac{1}{2} , \frac{d-2}{2d -2 } ; \frac{3d -4}{2d -2} ; - \frac{\rho^2}{L_2^{2(d-1)}} r^{- 2 (d-1)}\right) + \mu \ , \nonumber\\
\end{eqnarray}
where, again $\rho$ and $C_2$ are integration constants. Also, $\alpha_d$ is a constant that we haven't specified yet. In the above, $\mu$ is a constant such that the boundary condition $A_t(r_{\rm H}) = 0 $ is satisfied. It is easy to check that, asymptotically, the gauge field behaves as:
\begin{eqnarray}
A_t(r) = \mu + \beta_d \frac{\rho}{r^{d-2}} + \ldots \ ,
\end{eqnarray}
where $\beta_d$ is an overall constant. The dimensionless parameter (in our choice of units), $\rho$, captures the presence of a non-vanishing flavour density.

The temperature is given by
\begin{eqnarray}
T & = & \left. \frac{1}{4\pi} \frac{\partial_r g_{tt}}{\sqrt{- g_{tt} g_{rr}}} \right|_{r_{\rm H}} \nonumber\\
& = & \frac{ r_{\rm H}}{4 \pi  L^2} \left(d L_2^2 - \frac{2  \epsilon}{d-1}   \frac{L^2}{\left(L_2 r_{\rm H}\right)^{d+1}} \sqrt{ \rho^2 + ( L_2 r_{\rm H})^{2 (d-1)}}\right)  \ .
\end{eqnarray}
The extremal case is thus described by
\begin{eqnarray}
\frac{2  \epsilon}{d-1} \frac{L^2}{\left(L_2 r_{\rm H}\right)^{d+1}} \sqrt{ \rho^2 + ( L_2 r_{\rm H})^{2 (d-1)}} = d L_2^2 \ .
\end{eqnarray}
Imposing the boundary condition that $f(r) \to 1$ as $r \to 1$, one gets:
\begin{eqnarray}
L_2^2 = \frac{d \left( d -1 \right) L^2}{d \left( d -1 \right) - 2 L^2 \epsilon} \ .
\end{eqnarray}
In the extremal case, the line element finally takes the form:
\begin{eqnarray}
&& ds^2 = L_2^2 \left( - \ell^2 r^2 dt^2 + \frac{dr^2}{\ell^2 r^2 }  + r_{\rm H}^2 d\vec{x}^2 \right) \ ,  \label{ads2_gen1} \\
&& \ell^2 = \frac{(d-1) L_2^2}{2 d L^2} \left(d^2 - \frac{4 L^4 \epsilon^2}{ \left( d -1 \right)^2} \right) \ . \label{ads2_gen2}
\end{eqnarray}
In the extremal background, now, one can analyze the scalar field equation of motion to obtain the following power-law behaviours:
\begin{eqnarray}
&& \phi = A r^{- \frac{1}{2} - \nu_k} + B r^{- \frac{1}{2} + \nu_k} \ , \\
&& \nu_k = \sqrt{\frac{1}{4} + \frac{1}{\ell^2} \left( m^2 L_2^2 + \frac{k^2}{r_{\rm H}^2} \right) } \ . \label{nukgen}
\end{eqnarray}
The exponent $\nu_k$ becoming complex-valued is a sign of instability\cite{Hartnoll:2016apf}.

From (\ref{nukgen}), one can easily calculate the corresponding BF-bound at vanishing momentum (setting $k=0$), which is given by
\begin{eqnarray}
m^2 L^2 \ge - \frac{d(d-1)}{8} + \frac{\epsilon^2}{d(d-1)} \ . 
\end{eqnarray}
It is clear from the expression above, that, a non-vanishing $\epsilon$, lifts the BF-bound towards a positive value and is able to less tolerate negative mass-squared values. In fact, setting $\epsilon= \epsilon_{\rm max} = d(d-1)/2$, we get $m^2 L^2 \ge 0$ to ensure stability.\footnote{This is expected, since $\epsilon = \epsilon_{\rm max}$ yields $\Lambda_{\rm eff} = 0$, which corresponds to asymptotically flat geometry. Thus, one should recover the usual tachyonic bound. }

On the other hand, we can certainly ask whether the infra-red, which is non-perturbative in $\epsilon$\cite{Kundu:2016oxg}, facilitates a BF-bound violation, compared to the UV flavoured-CFT. The UV stability regime is: $m^2 L_2^2 \ge - d^2 /4$, while the IR BF-bound is: $m^2 \left(L_2/ \ell \right)^2 \ge - 1 / 4$. To check the pattern, let us begin with the BF-bound saturating mass in the IR, namely $m^2 \left(L_2 /\ell \right)^2 = - 1/4$, which corresponds to:
\begin{eqnarray}
m^2 L_2^2 = - \frac{\ell^2}{4} = - \frac{d^2}{8} + \left( \frac{d}{8} - \frac{\epsilon}{4} \right) \ .
\end{eqnarray}
It is straightforward to observe now, that, the IR BF-bound saturating mass is safely above the UV BF-bound, and thus a mode that is stable from an UV perspective, may become unstable in the IR.

Before moving further, let us isolate the $d=4$ case, since we do not loose the generic features for this case. Furthermore, to explore the effect of the Chern-Simons coefficient, we need to specifically focus on this particular case. The exact charged solution is give by
\begin{eqnarray}
&& ds^2 =  L_2^2 \left( - r^2 f(r) dt^2 + r^2 d\vec{x}^2 + \frac{dr^2}{r^2 f(r)} \right) \ , \label{exactsoln} \\
&& f(r) = \frac{L_2^2}{L^2} - \frac{2 \rho \epsilon}{3 L_2} \frac{1}{r^3} \, _2F_1 \left(-\frac{1}{2},\frac{1}{6};\frac{7}{6};-\frac{L_2^6 r^6}{\rho^2}\right) + \frac{C_1}{r^4} \ , \\
&& F = - \frac{\rho L_2^2  }{\sqrt{\rho^2 + \left( L_2 r \right)^6}} dt \wedge dr = - A_t' dt \wedge dr \ , \quad {\rm with} \quad L_2^2 = \frac{6 L^2 }{ 6 - \epsilon L^2 } \ .
\end{eqnarray}
The chemical potential is obtained by integrating $F$ from the horizon to infinity, which yields:
\begin{eqnarray}
\mu = \int_{r_{\rm H}}^{\infty} A_t' dr = \frac{ L_2 \Gamma \left( \frac{1}{3} \right) \Gamma \left(\frac{7}{6}\right) \rho  ^{1/3}}{\sqrt{\pi } } - \left( L_2^2 r_{\rm H} \right)   \, _2F_1 \left( \frac{1}{6}, \frac{1}{2}; \frac{7}{6};-\frac{L_2^6 r_{\rm H}^6}{ \rho^2} \right) \ .
\end{eqnarray}
Here $_2F_1$ is the hyergeometric function. The corresponding temperature is given by
\begin{eqnarray}
T  = - \frac{r_{\rm H}}{12 \pi} \left(- \frac{12 L_2^2}{L^2} + \frac{2  \epsilon  \sqrt{ L_2^6 r_{\rm H}^6 + \rho^2}}{ L_2 r_{\rm H}^3}  \right) \ . 
\end{eqnarray}
The extremal limit corresponds to setting $T=0$, which yields:
\begin{eqnarray}
r_{\rm H}^3 = \frac{\rho L^2 \epsilon }{ L_2^3 \sqrt{36 - L^4 \epsilon ^2}} \ .
\end{eqnarray}
In the extremal case, the line element in (\ref{exactsoln}) takes the form:
\begin{eqnarray}
&& ds^2 = L_2^2 \left( - \ell^2 r^2 dt^2 + \frac{dr^2}{\ell^2 r^2 }  + r_{\rm H}^2 d\vec{x}^2 \right) \  , \quad  \ell^2 = \frac{ L_2^2 \left( 36 - L^4 \epsilon ^2\right)}{6 L^2} \ ,  \label{ads2dim51} \\
&& A_t (r) = \frac{3 L_2^2 \Gamma \left(\frac{4}{3}\right)}{\Gamma \left(\frac{1}{3}\right)} r + \mu \ ,  \quad {\rm with} \quad \mu = \lim_{r\to 0}A_t (r) = \frac{\sqrt[3]{\rho} L_2 \Gamma \left(\frac{1}{6}\right) \Gamma \left(\frac{4}{3}\right)}{2 \sqrt{\pi }}    \ .  \label{ads2dim52}
\end{eqnarray}
Now, one can analyze, for example, a Klein-Gordon field and derive a deformed BF-bound.

So far, we have only discussed a possible condensation of neutral scalar fields, within the realm of possible BF-bound violation. One interesting, and rather extensively studied instability is due to the condensation of charged scalars in an extremal AdS$_2$-region, which leads to the models of holographic superconductors\cite{Gubser:2008px, Hartnoll:2008vx, Horowitz:2009ij}. One typically studies the following action:
\begin{eqnarray}
S = S_{\rm background} + \beta \int d^{d+1} x \sqrt{-g} \left( - \left| \partial_\mu \psi - i Q A_\mu \psi \right|^2 - M^2 \left| \psi \right|^2 \right) \ ,
\end{eqnarray}
where $S_{\rm background}$ is the action that is extremized on the AdS$_2$-geometry described above, $\beta$ is a coupling, which when tuned to take small values, decouples from the gravitational action and the corresponding sector is adequately described in the probe limit. This probe sector consists of a charged (complex) scalar, with charge $Q$ and mass $M$. Both the charge and the mass are {\it a priori} free parameters, which can only be fixed by a stringy embedding or consistency conditions, if any. The resulting Klein-Gordon equation takes the following form:
\begin{eqnarray}
&& \frac{1}{\sqrt{ |g |}} \partial_r \left( \sqrt{ |g |} g^{rr} \partial_r \psi \right) - M_{\rm eff}^2 \psi = 0 \ , \quad  M_{\rm eff}^2 = M^2 - Q^2 |g^{tt}| A_t A_t \ , \label{KGCharged}
\end{eqnarray}
where the equation above is obtained in the background (\ref{ads2_gen1})-(\ref{ads2_gen2}).

With the above, the equation in (\ref{KGCharged}) now takes the form:
\begin{eqnarray}
\partial_r^2 \psi + \frac{2}{r} \partial_r \psi - \frac{M_{\rm eff}^2 }{r^2} \psi = 0 \ , \quad {\rm with} \quad M_{\rm eff}^2 = \frac{1}{\ell^2} \left( M^2 L_2^2 - Q^2  \left( \frac{3 L_2^2 \Gamma \left(\frac{4}{3}\right)}{\Gamma \left(\frac{1}{3}\right)} \right)^2 \right) \ .
\end{eqnarray}
Now saturation of the AdS$_2$ BF-bound implies $M_{\rm eff}^2 = - 1/4$, which yields:
\begin{eqnarray}
M_{\rm IR, min}^2  = \frac{54 Q^2}{ \left( 36 - \epsilon^2 \right) } \left(  \frac{\Gamma \left(\frac{4}{3}\right)}{\Gamma \left(\frac{1}{3}\right)} \right)^2 - \frac{36 - \epsilon^2}{24} \ .
\end{eqnarray}
It is straightforward to check now, that, the right hand side above is a monotonically increasing function of $\epsilon$, for any given $Q$. Thus, for a fixed value of $Q$, however obtained, increasing $\epsilon$ raises the BF-bound, thereby facilitating a BF-bound violating instability. It is likely, therefore, that this instability may source the formation of a charged condensate, breaking a U(1) global symmetry in the putative dual field theory. Thus, a preliminary analysis hints that the presence of a flavour-sector, thus-far modelled by a simple DBI-matter field, catalyzes a superconductor/superfluid instability.

\section{Spatially Modulated Instability}

Now that we have observed a potential superconducting-type (or, to be precise, superfluid-type) instability, we want to explore, along similar directions, how the presence of flavours may affect a spatially modulated instability that typically forms in the large $N_c$ gauge theories with holographic duals\cite{Nakamura:2009tf, Donos:2011bh, Donos:2011qt}. Specifically, we study fluctuations within the action in (\ref{actm1})-(\ref{actm4}), in $d=4$ and around the solution in (\ref{ads2dim51})-(\ref{ads2dim52}). We will follow the same treatment in \cite{Nakamura:2009tf} (see also \cite{Liu:2016hqb}). We write the metric components and the gauge field strengths as:
\begin{eqnarray}
\tilde{g}_{\mu\nu} = g_{\mu\nu} + h_{\mu\nu} \ , \quad \tilde{A}_\mu = A_\mu + a_\mu \ , 
\end{eqnarray}
where $\{g_{\mu\nu}, A_\mu\}$ correspond to the background geometric data and $\{h_{\mu\nu}, a_\mu\}$ correspond to the fluctuations around it. For convenience, we will choose the gauge $h_{\mu r} = 0$ and $a_r = 0$. To further simplify our notation, we use $I, J, \ldots $ to denote $\{t, r\}$ coordinates (the AdS$_2$), $i, j, \ldots$ to denote $\{x^2, x^3\}$ coordinates. Furthermore, we consider the fluctuations in the basis: $e^{- i \omega t + i q y}$, where $y \equiv x^1$. Here, a non-vanishing $q$ represents spatial modulation. The fluctuations now can be classified according to the representations of the remaining O$(2)$ symmetry in the $\{x^2, x^3\}$-plane, as scalars, vectors and tensors. This is discussed in details in appendix A.

The simplest sector is the tensor one. In this sector, the fluctuations equations are given by (see, (\ref{teneqn3}))
\begin{eqnarray}
&& \nabla_{{\rm AdS}_2}^2 h_{ij} - M_{\rm eff}^2 h_{ij} = 0 \ , \quad M_{\rm eff}^2 =\frac{q^2}{L_3^2} \ . \label{tenmain2}
\end{eqnarray}
As discussed in appendix A, this mode does not violate the BF-bound.

Let us now consider the vector fluctuations, in which the Chern-Simons coupling plays an important role, as was observed in \cite{Nakamura:2009tf}. The equations of motion can be written as (see, the discussion in (\ref{vecteqn1})-(\ref{vecteqn6})):
\begin{eqnarray}
\nabla_{{\rm AdS}_2} \Psi_{\pm} = M_{\pm}^2 \Psi_{\pm} \ , \label{vectmaineqn4}
\end{eqnarray}
with
\begin{eqnarray}
&& \psi_i = \left( 2 \rho L_2^2 \epsilon \right)  a_i + L_3^3 \partial_r h_{ti} \ , \label{vectmaineqn1} \\
&& {\rm subsequently} \ , \quad \Psi_{+} = \left\{ a_2 + i a_3 , \psi_2 + i \psi_3 \right\} \ , \label{vectmaineqn2}  \\
&& {\rm and} \quad  \Psi_{-} = \left\{ a_2 - i a_3 , \psi_2 - i \psi_3 \right\} \ . \label{vectmaineqn3}
\end{eqnarray}
The corresponding mass matrices are given in (\ref{vecteqn5})-(\ref{vecteqn6}).

Given the above mass matrices, it is straightforward to obtain the eigenvalues, which we denote by $\lambda_{\pm}^{a}$, with $a=1,2$, respectively. The explicit expressions are somewhat cumbersome. Clearly, each eigenvalue is a function of the momentum $q$, and two dimensionless parameters: the flavour weight $\epsilon$ and the Chern-Simons coupling $\alpha$. To check the violation of BF-bound, we can proceed as follows: First, we minimize each eigenvalue with respect to the momentum $q$, by solving:
\begin{eqnarray}
&& \left. \partial_q \lambda_{\pm}^{a} \left(q, \epsilon, \alpha \right) \right|_{q = q_{\rm crit}}  = 0 \ , \\
&& {\rm and \, \, evaluate} \quad \lambda_{\pm}^{a} \left( q_{\rm crit}, \epsilon, \alpha \right)   = \lambda_{\pm}^{(0)a} \left(\epsilon, \alpha \right) \ .
\end{eqnarray}
Now, the condition of BF bound violation, in AdS$_2$, yields:
\begin{eqnarray}
\lambda_{\pm}^{(0)a}\left(\epsilon, \alpha \right) \frac{L_2^2}{\ell^2} < - \frac{1}{4} \ .
\end{eqnarray}
The critical saturation, $\lambda_{\pm}^{(0)a}\left(\epsilon, \alpha \right) = - \frac{1}{4} \frac{\ell^2}{L_2^2}$, defines a hyper-plane in the $\{\epsilon, \alpha \}$--space, that determines the stable regime. It is unwieldy to tackle this analytically, and instead we can proceed numerically.

It can be checked that the set $\left\{\alpha, \epsilon \right\}$ constitutes, as far as the BF-bound violating instability is concerned, competing parameters. Thus, we explore the competing effects in details now. Certain analytical insights are easily available in the limit $\epsilon \to 0$, in which we get:
\begin{eqnarray}
&& \lim_{\epsilon \to 0} \lambda_{\pm}^{(0)1} \frac{L_2^2}{\ell^2} = -\frac{2 \ 2^{2/3} A \sqrt[3]{\epsilon }}{\sqrt[3]{3} \rho^{2/3} q} + \frac{q^2}{\sqrt[3]{6} \rho^{2/3} \epsilon^{2/3}} + 2 + \O \left(\epsilon^{2/3} \right) \ldots \ , \\
&& \lim_{\epsilon \to 0} \lambda_{\pm}^{(0)2} \frac{L_2^2}{\ell^2} = \pm \frac{2\ 2^{2/3} A \sqrt[3]{\epsilon }}{\sqrt[3]{3} \rho^{2/3} q} \mp \frac{\sqrt[3]{2} A q}{3^{2/3} \rho^{4/3} \sqrt[3]{\epsilon }} + \O \left(\epsilon^{2/3} \right) + \ldots \ . \\
&& A = 2 \alpha \frac{\Gamma\left( \frac{3}{4} \right) }{\Gamma \left( \frac{1}{3} \right) } \ .
\end{eqnarray}
The above expressions clearly suggest that, with a non-vanishing $\epsilon$ and $A$ (and $A$ needs to be hold fixed), $\lambda_{\pm}^{(0)2}$ can become negative enough to violate the IR BF-bound. On the other hand, in the limit $\epsilon \to \epsilon_{\rm max}$ (here, $\epsilon_{\rm max} =6$, with $L=1$), we get:
\begin{eqnarray}
\lim_{\epsilon \to \epsilon_{\rm max}} \lambda_{\pm}^{(0)1, 2} \frac{L_2^2}{\ell^2} =  \frac{q^2}{2 \cdot 3^{1/3} \rho^{2/3}} \frac{1}{\left( \epsilon_{\rm max} - \epsilon \right)^{2/3}} + \O \left( \left( \epsilon_{\rm max} - \epsilon \right)^{-1/3} \right) \ldots \ , 
\end{eqnarray}
which, in the leading order, becomes strictly positive and hence hints towards stable modes. 
\begin{figure}[h!]
\centering
\includegraphics[scale=0.45]{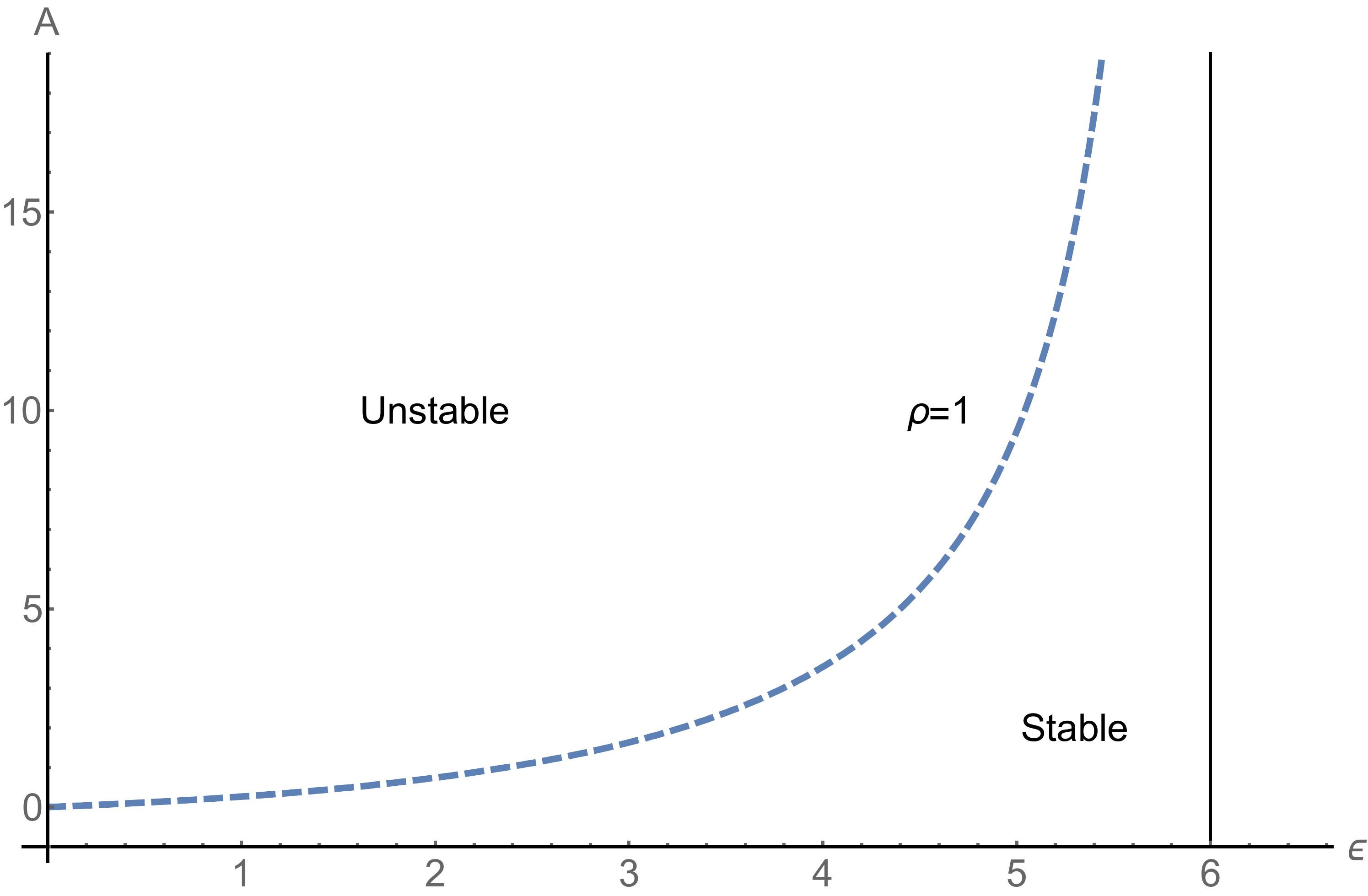}
\caption{\small We have shown the phase diagram in the $\{\epsilon, \alpha\}$-plane, setting $\rho = 1$. The black vertical line corresponds to $\epsilon = \epsilon_{\rm max}$. Stable and unstable regions are also marked. Near $\epsilon \to \epsilon_{\rm max}$, $A_{\rm crit} \sim \left( \epsilon_{\rm max} - \epsilon \right)^{-1/2}$.}
\label{alphaepsphase}
\end{figure}
One can now explore numerically. The result is shown in fig.~\ref{alphaepsphase}, which implies that, if we treat both the Chern-Simons coefficient and the flavour weight as free parameters in the theory\footnote{Changing the couplings in a Lagrangian corresponds to changing the theory, however, here we adopt the view point of exploring condensed-matter systems. In that case, given a Hamiltonian, one freely tunes the couplings to access various phases describable by the said Hamiltonian.}, the spatially modulated instability sets in for increasingly larger values of the Chern-Simons coefficient as the flavour weight is increased. Near $\epsilon \to \epsilon_{\rm max}$, this instability seems to disappear. Since any non-zero value of $\rho$ is physically equivalent modulo overall scaling, we have chosen $\rho = 1$.

One may now wonder about the robustness of the result, specially in the context of an appropriate UV-completion of the theory. We will now demonstrate that, an inherently IR observer who only knows the meaning of $\epsilon$ without the knowledge of the UV, observes essentially the same physics in the deep-IR limit. This lack of UV knowledge enables the IR-observer pick her/his measurement scale at will to describe the gravitational AdS$_2 \times {\mathbb R}^3$-geometry. One natural choice is in terms of the following exact solution of (\ref{ein}) and (\ref{max}):
\begin{eqnarray}
&& ds^2 = R^2 \left( -r^2 dt^2 + \frac{dr^2}{r^2} \right) + d\vec{x}^2 \ , \\
&& R^2 = \frac{6}{36 - \epsilon^2} \ , \quad F = \frac{2}{\sqrt{36 - \epsilon^2}} dt \wedge dr \ , \quad {\rm with} \quad \Lambda = - 6 \ .
\end{eqnarray}
Note that, as pointed out in \cite{Kundu:2016oxg}, in the above IR-description, the flux and the radius of curvature are locked together, and thus there is no independent concept of a density. Furthermore, the IR AdS$_2$ is a non-perturbative solution in $\epsilon$, since the precise $\epsilon = 0 $ limit does not admit this solution.

One can now proceed to analyze, specially the vector sector of the fluctuation modes, as described above. Below we only briefly present the results. It is straightforward to diagonalize the mass matrices. The corresponding eigenvalues $\lambda_{\pm}^a \left(\epsilon, \alpha \right)$ are again functions of two independent dimensionless variables. Now, in the limiting case of $\epsilon \to \epsilon_{\rm max} =6$, some analytical insights are easily available. The mass eigenvalues in this case yield:
\begin{eqnarray}
&& \lambda_{\pm}^{(0)1} R^2 = \frac{q^2}{2 \left( \epsilon_{\rm max} - \epsilon\right) } + \frac{q^2}{24} + \O \left (\left( \epsilon_{\rm max} - \epsilon \right) ^2 \right)   + \ldots \ , \\
&& \lambda_{\pm}^{(0)2} R^2 = \frac{q^2}{2 \left( \epsilon_{\rm max} - \epsilon\right) } +  \frac{ 16 \sqrt{3} \alpha q}{\sqrt{\epsilon_{\rm max} - \epsilon}} + \left( 2 - \frac{q^2}{8} \right) + \O \left (\left( \epsilon_{\rm max} - \epsilon \right) ^2 \right)   + \ldots \ .
\end{eqnarray}
Clearly, $\lambda_{\pm}^{(0)1} R^2$ cannot become negative. On the other hand, $\lambda_{\pm}^{(0)2} R^2$ can become negative and violate the AdS$_2$ BF bound for negative values of $q$, provided, $\alpha \sim \left( \epsilon_{\rm max} - \epsilon \right)^{-1/2}$. In the limit $\epsilon \to \epsilon_{\rm max}$, thus, the instability-inducing Chern-Simons coupling diverges.\footnote{The argument above should be viewed as a double expansion, one in $\epsilon$ and one in $\alpha$. It can be checked that relating the two expansions does not miss out any term, as far as our argument above is concerned.} 
\begin{figure}[h!]
\centering
\includegraphics[scale=0.5]{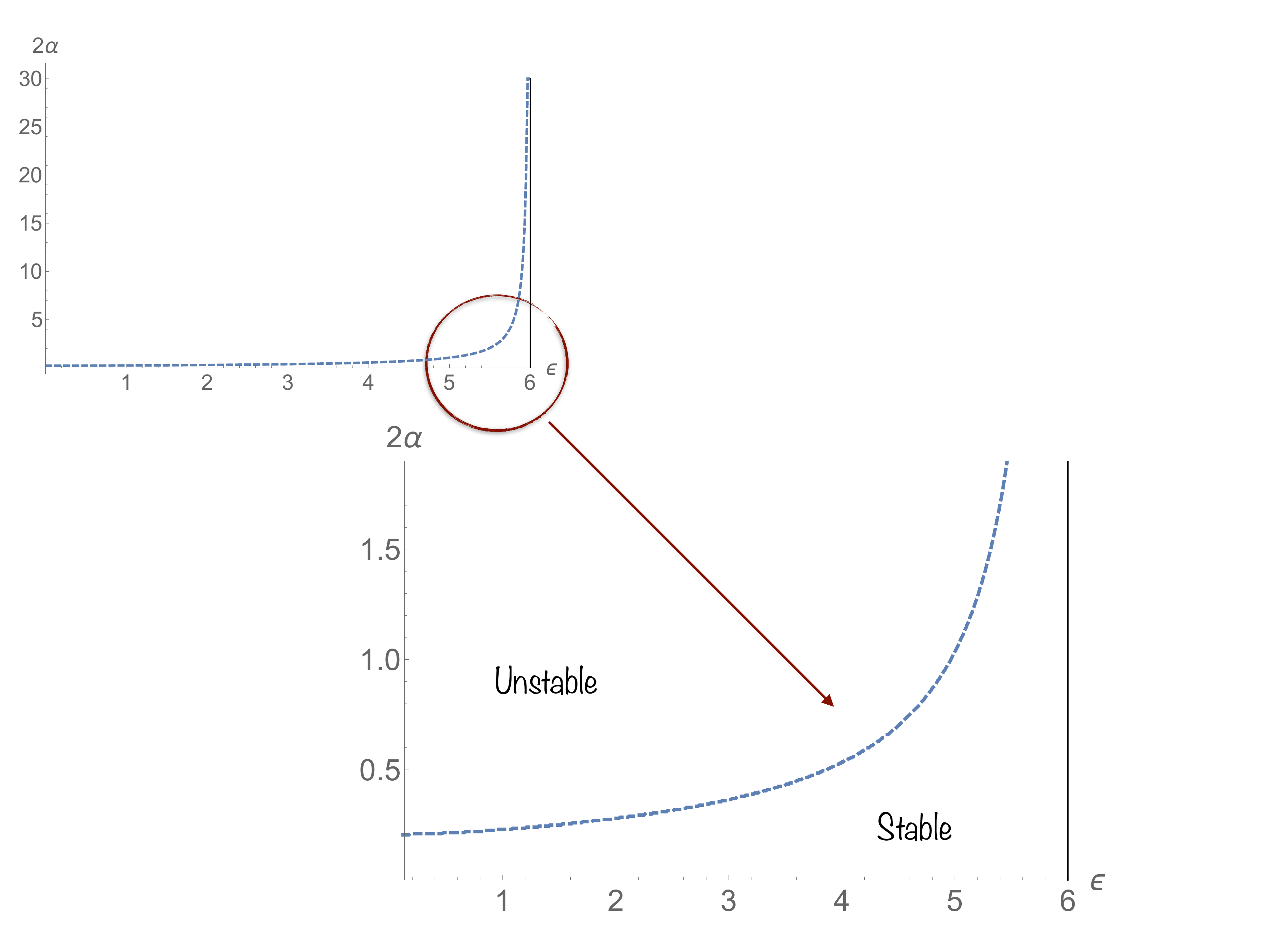}
\caption{\small We have shown the phase diagram in the $\{\epsilon, \alpha\}$-plane. The blue dashed curve is the pictorial representation of $\lambda_{\pm}^{(0)a}\left(\epsilon, \alpha \right) R^2 = - \frac{1}{4} $, the black vertical line corresponds to $\epsilon = \epsilon_{\rm max}$. Stable and unstable regions are also marked. The critical value of $\alpha$ diverges near $\epsilon \to \epsilon_{\rm max}$, as $\alpha_{\rm crit} \sim \left( \epsilon_{\rm max} - \epsilon \right)^{-1/2}$.}
\label{alphaepsphaseIR}
\end{figure}
We have presented our numerical results in fig.~\ref{alphaepsphaseIR} which confirms what we have already observed.

\section{Conclusions}

In this article, we have demonstrated that, in a simple {\it bottom up} or phenomenological model, introducing $N_f$ fundamental matter in an SU$(N_c)$-gauge theory in the limit $\O\left(N_f \right) \sim \O\left(N_c \right)$, appears to facilitate an instability similar to the formation of a superconducting/superfluid phase, and suppress a spontaneous breaking of translational symmetry as $\left(N_f / N_c \right)$ is tuned. This is, in spirit although not precisely, similar to the conclusion in \cite{Shuster:1999tn}.

There are many potential limitations of our work. First, to draw any precise conclusion in comparison with \cite{Shuster:1999tn}, one needs a stringy/supergravity construction where all fields in the bulk are, in principle, identifiable with a precise boundary theory operator. Towards that, the models studied in \cite{Faedo:2014ana, Faedo:2015urf, Faedo:2016jbd, Faedo:2016cih, Faedo:2017aoe} may be natural to explore the fate of such instabilities further, and perhaps more of the corresponding phase diagram and equations of state. See, {\it e.g.}~\cite{Bigazzi:2013jqa} for early analyses on a {\it top down} model and \cite{Faedo:2017aoe} for recent progress towards this.

Going back to the model at hand, because of the simplicity, it deserves further study. For example, the candidate ground state ({\it i.e.}~infra-red) solutions of the action in the presence of both density and a magnetic field raises interesting possibilities\cite{InProgress}, similar to the studies for a closely related system in \cite{DHoker:2009ixq}. A potentially intriguing issue is to explore how much of what we observed here survives in the myriad of candidate ground states.

The presence of a magnetic field is specially interesting. In holographic models, it catalyzes spontaneous breaking of symmetry\cite{Filev:2007gb, Bergman:2008sg, Johnson:2008vna, Filev:2010pm, Filev:2011mt, Alam:2012fw} by reducing the effective dimensionality of the system. In the context of \cite{Kundu:2016oxg}, the corresponding IR is an AdS$_3$ which is dual to a $2$-dim CFT. The central charge of the CFT, which in turn is related to the curvature scale of the AdS$_3$, is expected to monotonically decrease along the flow from the UV $4$-dim CFT. Thus, it is reasonable to expect that the BF-bound violation would be facilitated along the RG-flow. Interestingly, this seems to hint a possible connection between the existence of a $c$-theorem involving flows from a $4$-dim CFT to an effectively $2$-dim CFT and the role of fundamental matter in inducing/impeding various instabilities. In this article, we take the first preliminary step towards exploring these issues further.

\section{Acknowledgements}

AK would like to thank A.~Faedo, N.~Kundu, D.~Mateos, C.~Pantelidou, J.~Tarr\'{i}o for collaborations on related topics. AK would further like to thank Centro de Ciencias de Benasque Pedro Pascual during the ``Gravity-New perspective from strings and higher dimensions" workshop, for providing a wonderful and stimulating environment during the final stages of this work. Finally, we humbly acknowledge the support from the people of India, towards research in basic sciences.

\appendix

\section{Computing Fluctuations}

In this appendix, we present covariant formulae for calculating the fluctuations of the action described in this article. For completeness, we include here the details, including some that are already discussed in the main text, focussing on $d=4$. The action is:
\begin{eqnarray}
&& S_{\rm full} = S_{\rm gravity} + S_{{\rm DBI} + {\rm CS}} \ , \label{act1} \\
&& S_{\rm gravity} = \frac{1}{2 \kappa^2} \int d^{5} x \sqrt{- {\rm det} g} \  \left(R - 2 \Lambda \right) \ , \label{act2} \\
&& S_{{\rm DBI} + {\rm CS}} = - \tau \int d^{5} x \  \sqrt{- {\rm det}  \left( g + F \right) } + \frac{\alpha\tau}{3!} \int A\wedge F \wedge F \ . \label{act3}
\end{eqnarray}

The equations of motion are:
\begin{eqnarray}
&& R_{\mu\nu} - \frac{1}{2} R g_{\mu\nu} + \Lambda g_{\mu\nu} = T_{\mu\nu} \ , \label{classsol1} \\
&& {\rm with} \quad T_{\mu\nu} = - \left( \kappa^2 \tau \right)  \frac{\sqrt{ - \left(g + F \right)}}{\sqrt{- g}} \S_{\mu\nu}  \ , \quad \S_{\mu\nu} = g_{\mu\alpha} \S^{\alpha\beta} g_{\beta \nu} \ , \label{classsol2} \\
&& {\rm and} \quad \partial_\mu  \left[ \sqrt{- \left(g + F \right) } \A^{ \mu\nu} \right] + \frac{\alpha}{2} \epsilon^{\nu\alpha\beta\rho\sigma} F_{\alpha\beta} F_{\rho\sigma} = 0 \ . \label{classsol3}
\end{eqnarray}
To find the equations for fluctuations, we write
\begin{eqnarray}
\tilde{g}_{\mu\nu} = g_{\mu\nu} + \delta g_{\mu\nu}  \ , \quad \tilde{F}_{\mu\nu} = F_{\mu\nu} + \delta F_{\mu\nu}  \ .
\end{eqnarray}
At the leading order, we have
\begin{eqnarray}
&& \tilde{R}_{\mu\nu} = R_{\mu\nu}  + \delta R_{\mu\nu}  \ , \\
&& \delta R_{\mu\nu} = \frac{1}{2} \left( - \nabla^2 \delta g_{\mu\nu} - \nabla_\mu \nabla_\nu \delta g + \nabla^\sigma \nabla_\nu \delta g_{\sigma\mu} + \nabla^\sigma \nabla_\mu \delta g_{\sigma\nu} \right) \ , \\
&& {\rm with} \quad \delta g = \delta g_{\mu\nu} g^{ \mu\nu} \ .
\end{eqnarray}
And also
\begin{eqnarray}
\tilde{R} & = & R + \delta R \ , \label{delR1} \\
\delta R & = & \delta R_{\mu\nu} g^{\mu\nu} + R_{\mu\nu} \delta g^{\mu\nu } \nonumber\\
& = & - \nabla^2 \delta g + \nabla_\mu\nabla_\nu \delta g^{\mu\nu} + R_{\mu\nu} \delta g^{\mu\nu } \ .  \label{delR2}
\end{eqnarray}
Now, to calculate the variations in $\S$, $\A$, $T_{\mu\nu}$ and $\sqrt{- (g+F)}$, let us note that
\begin{eqnarray}
\sqrt{- X + \delta X} \approx \sqrt{ - X} \left[ 1 + \frac{1}{2} {\rm Tr} \left( X^{-1} \delta X \right) + \ldots \right] \ .
\end{eqnarray}
Thus we get
\begin{eqnarray}\label{detexpand}
&& \sqrt{- \left( g + F +  \delta g + \delta F \right) } = \sqrt{- \cM_+} \left[ 1 + \frac{1}{2} {\rm Tr} \left( \cM_{+}^{-1} \delta\cM_+ \right) \right] \ , \\
&& \cM_\pm = g \pm F \ , \quad \delta\cM_\pm = \delta g \pm \delta F \ .
\end{eqnarray}
To proceed further, we will treat various quantities as matrices, avoiding writing down indices. Now, to further determine the contributions coming from $\S$, $\A$ and $T_{\mu\nu}$, let us first note that
\begin{eqnarray}
\left( X + \delta X \right) ^{-1 } = X^{-1} - X^{-1} \delta X X^{-1} \ .
\end{eqnarray}
Now we can write:
\begin{eqnarray}
 \S & = & \cM_+^{-1} g \cM_-^{-1} + \delta \S \ , \\
 \delta \S & = & \cM_+^{-1} \delta g \cM_-^{-1} - \cM_+^{-1} \delta \cM_+ \cM_+^{-1} g \cM_-^{-1} - \cM_+^{-1} g \cM_-^{-1} \delta \cM_- \cM_-^{-1} \ . \\
 \A & = & - \cM_+^{-1} F \cM_-^{-1} + \delta \A \ , \\
 \delta \A & = & - \cM_+^{-1} \delta F \cM_-^{-1} + \cM_+^{-1} \delta \cM_+ \cM_+^{-1} F \cM_-^{-1} + \cM_+^{-1} F \cM_-^{-1} \delta \cM_- \cM_-^{-1} \ . \nonumber\\
\end{eqnarray}
In the above expressions, the products are all matrix products. Now, the only quantity left to calculate is the fluctuation in the energy-momentum tensor. This is evaluated to be:
\begin{eqnarray}
\delta T^{\mu\nu} = - \left(\kappa^2 \tau \right) \frac{\sqrt{-\cM_+}}{\sqrt{- g}} \left[ \delta \S + \frac{1}{2} \S \left( {\rm Tr} \left( \cM_+^{-1} \delta \cM_+\right) - {\rm Tr} \left( g^{-1} \delta g\right) \right) \right]^{\mu\nu} \ .
\end{eqnarray}
The energy-momentum tensor, with indices lowered, is given by
\begin{eqnarray}
\delta T_{\mu\nu} = g_{\mu\alpha} \delta T^{\alpha\beta} g_{\nu\beta} + \delta g_{\mu\alpha} T^{\alpha\beta} g_{\nu\beta} +  g_{\mu\alpha} T^{\alpha\beta} \delta g_{\nu\beta} \ .
\end{eqnarray}

Using the results above, the gauge field equation of motion, at the leading order in fluctuations around the classical vacua, takes the following form:
\begin{eqnarray}
\partial_\mu \left[\sqrt{- {\rm det} \cM_+} \ \delta \A^{\mu\nu} \right] & + & \frac{1}{2} \partial_\mu \left[ \sqrt{- {\rm det} \cM_+} \ {\rm Tr} \left( \cM_+^{-1} \delta\cM_+\right) \A^{\mu\nu}\right] \nonumber\\
& + & \alpha \epsilon^{\nu\alpha\beta\rho\sigma} F_{\alpha\beta} \delta F_{\rho\sigma} = 0  \ .
\end{eqnarray}
The Einstein equation can be written as:
\begin{eqnarray}
\delta R_{\mu\nu} - \frac{1}{2} \left( R - 2 \Lambda \right) \delta g_{\mu\nu} - \frac{1}{2} \delta R g_{\mu\nu}  = \delta T_{\mu\nu} \ .
\end{eqnarray}
We still have gauges to fix. Henceforth, we will rewrite: $\delta g \equiv h$ and $\delta F = f$.

A particularly useful gauge is to choose: $h_{\mu r} = 0$ and $a_r = 0$. Suppose, further, that we only consider the fluctuations to propagate only along $x^1$-direction: $e^{- i \omega t + i q x^1} = e^{- i \omega t + i q y}$, with $y\equiv x^1$. Thus the O$(2)$-symmetry in the $\{x^2, x^3\}$-plane remains unbroken. We can now classify the fluctuations, as representations of this unbroken O$(2)$-symmetry, in the following categories:
\begin{eqnarray}
&& {\rm spin} \, 0: \left\{h_{tt}, h_{yy}, h_{ty}, f_{tr}, f_{ty}, f_{yr} \right\} \ , \\
&& {\rm spin} \, 1:   \left\{h_{t x^2}, h_{t x^3}, h_{y x^2}, h_{y x^3}, f_{tx^2}, f_{tx^3}, f_{yx^2}, f_{yx^3}, f_{r x^2}, f_{r x^3} \right\} \ , \\
&& {\rm spin} \, 2:   \left\{h_{x^2 x^2}, h_{x^3 x^3}, h_{x^2 x^3}, f_{x^2 x^3} \right\} \ .
\end{eqnarray}
We will only explicitly write down the equations for the tensor and the vector sector, but not the scalar sector.

Note that, we are working in the background: 
\begin{eqnarray}
&& ds^2 = L_2^2 \left( - \ell^2 r^2  dt^2 + \frac{dr^2}{\ell^2 r^2} + r_{\rm H}^2 d\vec{x}^2 \right) \ , \quad \ell^2 = \frac{L_2^2}{6L^2} \left( 36 - L^4 \epsilon^2 \right) \ , \\
&& L_2^2 = \frac{6 L^2}{6 - L^2 \epsilon} \ , \quad r_{\rm H}^3 = \frac{\rho L^2 \epsilon}{L_2^3 \sqrt{36 - L^4 \epsilon^2}} \ ,\quad  A_t (r) =  3 L_2^2 r \frac{\Gamma\left( \frac{3}{4}\right) }{\Gamma\left( \frac{1}{3} \right) } \ , \quad \frac{L_3^2}{L_2^2} = r_{\rm H}^2 \ . \nonumber\\
\end{eqnarray}
A further useful fact is the following:
\begin{eqnarray}
\nabla^2 \varphi = \left( \nabla_{{\rm AdS}_2}^2 + \nabla_{{\mathbb{R}^3}}^2 \right) \varphi \ .
\end{eqnarray}
Here $\varphi$ is a generic field with arbitrary spin. Also observe that
\begin{eqnarray}
\nabla_{{\rm AdS}_2} \varphi = \frac{\ell^2}{L_2^2} \left[ \frac{\omega^2}{\ell^4 r^2} + \partial_r \left( r^2 \partial_r\right)  \right] \varphi \ ,
\end{eqnarray}
where $\varphi$ is a generic field, written in the basis of $e^{-i \omega t + i q y}$.

\subsection{Tensor sector}

The simplest sector to look at is the tensor one. We can consistently set $f_{x^2x^3} = 0$, which imposes $h_{x^2x^2} + h_{x^3x^3} = 0$, {\it via} Maxwell's equations. This condition is also intuitive, since the trace part of the $h_{\mu\nu}$-fluctuations belong to the scalar sector, which has been set to zero. The fluctuations equations, in this sector, are given by
\begin{eqnarray}
&& \nabla_{{\rm AdS}_2}^2 h_{ij} - M_{\rm eff}^2 h_{ij} = 0 \ , \quad M_{\rm eff}^2 =\frac{q^2}{L_3^2} \ . \label{teneqn3}
\end{eqnarray}
In the above $i, j = 2, 3$. Clearly, the spatial momentum contributes positively towards the mass. In the above, we have calculated $M_{\rm eff}^2$ on-shell. The corresponding BF-bound is simply given by:
 $M_{\rm eff}^2 \ge - \frac{1}{4} \frac{\ell^2}{L_2^2}$, which is trivially satisfied for real-valued momentum.

\subsection{Vector sector}


Let us now look at the vector sector. First, we fix the orientation of the background manifold, by setting $\varepsilon_{tr} = 1$, $\varepsilon_{yx^2x^3}=1$, while the Levi-Civita tensor is defined as: $\epsilon_{\nu\alpha\beta\rho\sigma} = \sqrt{-{\rm det} g} \ \varepsilon_{\nu\alpha\beta\rho\sigma}$. The Maxwell's equations, by direct calculation, yield:
\begin{eqnarray}
&&  L_3 L_2^2 \left(\rho^2 + L_3^6 \right) \frac{\ell^2}{L_2^2} \left(  \partial_r \left(   r^2 a_{2/3}'(r) \right) + \frac{\omega ^2 a_{2/3}(r)}{\ell^4 r^2}\right) \nonumber\\
&& +  \left( - L_2^2 L_3^5 q^2 a_{2/3} (r) + \sqrt{ \rho^2 + L_3^6} \left( \rho L_3 h_{t2/3}'(r) \pm 2 i A q a_{3/2}(r) \right) \right) = 0 \ . \label{eqt1}
\end{eqnarray}
On the other hand, Einstein's equations give:
\begin{eqnarray}
&& {t i}: \ell^2 r^2 \left( - \frac{2 \rho \epsilon a_i'(r)}{L_3^3} - \frac{ h_{ti}''(r)}{ L_2^2}\right) + h_{ti}(r) \left(\frac{2 L_3^3 \epsilon }{\sqrt{\rho^2 + L_3^6}} + \frac{2 \ell^2}{L_2^2} - \frac{12}{L^2} + \frac{q^2}{L_3^2}\right) + \frac{q \omega  h_{yi}(r)}{L_3^2} = 0 \ , \nonumber\\
&& i = 2, 3 \ , \quad y \equiv  x^1 \ . \label{eqt2}
\end{eqnarray}
\begin{eqnarray}
&& {r i}: 2 \rho L_2^2 \omega  \epsilon  a_i(r) + \ell^2 L_2^2 L_3 q r^2 h_{yi}'(r) + L_3^3 \omega  h_{ti}'(r) = 0 \ , \label{eqt3} \\
&& {yi} : \frac{2 L_3^3 \epsilon  h_{yi}(r)}{\sqrt{\rho^2 + L_3^6}} + 2 h_{yi}(r) \left(\frac{\ell^2}{L_2^2 } - \frac{6}{L^2} \right) - \frac{\ell^2}{L_2^2} \left( \partial_r \left( r^2 h_{yi}' \right) + \frac{\omega ^2}{\ell^4 r^2 } h_{yi}(r)\right) + \frac{q \omega h_{ti}(r) }{\ell^2 L_2^2 r^2} = 0 \ . \nonumber\\ \label{eqt4}
\end{eqnarray}

Using (\ref{eqt3}) and (\ref{eqt2}), we can derive
\begin{eqnarray}
\frac{\ell^2 L_2^2}{L_3^2} \frac{q}{\omega} r^2 \nabla_{{\rm AdS}_2}^2 h_{yi} + \left( \frac{2 L_3^3 \epsilon }{\sqrt{ \rho^2 + L_3^6}} + \frac{2 \ell^2}{ L_2^2} - \frac{12}{L^2} + \frac{q^2}{L_3^2} \right) h_{ti} = 0 \ ,
\end{eqnarray}
We can now rewrite the system of equations as follows:
\begin{eqnarray}
&& \frac{L_3^6 + \rho^2 }{L_3^4}  \nabla_{\rm AdS_2}^2 a_{2/3} - q^2 a_{2/3} + \frac{\rho \sqrt{ L_3^6 + \rho^2}}{L_2^2 L_3^4} \partial_r h_{t2/3} \pm  i \left( \frac{2 A q }{ L_2^2 L_3^5} \right) \sqrt{L_3^6 + \rho^2} \  a_{3/2} = 0  \ , \nonumber\\ \\
&& \nabla_{\rm AdS_2}^2 h_{yi} + \left( \frac{12}{L^2} - \frac{2\ell^2}{L_2^2} - \frac{2\epsilon L_3^3}{\sqrt{L_3^6 + \rho^2 }} \right) h_{yi} + \omega q \left(\frac{h_{ti}}{r^2\ell^2 L_2^2} \right) = 0 \ , \\
&& \nabla_{\rm AdS_2}^2 h_{yi} + \frac{\omega}{q} \left( \frac{2 L_3^3 \epsilon }{\sqrt{ \rho^2 + L_3^6}} + \frac{2 \ell^2}{ L_2^2} - \frac{12}{L^2} + \frac{q^2}{L_3^2}  \right) \frac{L_3^2}{L_2^2} \frac{h_{ti}}{r^2 \ell^2} = 0 \ .
\end{eqnarray}
The last couple of equations are identical, when evaluated on-shell.

It is easy to demonstrate that, the above Einstein equations in (\ref{eqt2})-(\ref{eqt4}) are not independent. This is best shown by establishing the following identity:
\begin{eqnarray}
\gamma_1 ({\rm \ref{eqt2}}) + \gamma_2  ({\rm \ref{eqt4}}) + \gamma_3 \left[ \partial_r \left({\rm \ref{eqt3}} \right) \right] + \gamma_4 \left({\rm \ref{eqt3}} \right) = 0 \ , \label{constr2}
\end{eqnarray}
where $\gamma_i$ are constants that are determined by $\epsilon$, $\ell$ and $\rho$. Thus the constraint in (\ref{constr2}) demonstrates it is sufficient to consider only two Einstein equations, which are independent. We now need to construct a convenient combination of the fluctuation fields that will diagonalize the equations of motion.

To that end, we define the following:
\begin{eqnarray}
&& \psi_i = \left( 2 \rho L_2^2 \epsilon \right)  a_i + L_3^3 \partial_r h_{ti} \ , \label{vecteqn1} \\
&& {\rm subsequently} \ , \quad \Psi_{+} = \left\{ a_2 + i a_3 , \psi_2 + i \psi_3 \right\} \ , \label{vecteqn2}  \\
&& {\rm and} \quad  \Psi_{-} = \left\{ a_2 - i a_3 , \psi_2 - i \psi_3 \right\} \ . \label{vecteqn3}
\end{eqnarray}
With these redefinitions, the equations of motion in (\ref{eqt1})-(\ref{eqt4}) can be written as:
\begin{eqnarray}
\nabla_{\rm AdS_2} \Psi_{\pm} = M_{\pm}^2 \Psi_{\pm} \ , \label{vecteqn4}
\end{eqnarray}
where $M_{\pm}$ is the corresponding mass-matrix, which is given below:
\begin{eqnarray}
M_+ = \left(
\begin{array}{cc}
 \frac{12 \epsilon  \left(36 - \epsilon ^2\right) \rho^2 + q^2 \epsilon ^2 \sqrt[3]{\rho \epsilon  \left(36 - \epsilon ^2\right)} \rho - 2 A q (6 - \epsilon ) \epsilon ^{2/3} \left(36 - \epsilon ^2\right)^{2/3} \rho^{2/3}}{36 \rho^2 \epsilon } & -\frac{(6 - \epsilon )^2 (\epsilon + 6)}{36 \rho \epsilon } \nonumber\\
 -\frac{12 q^2 \sqrt[3]{\rho \epsilon  \left(36 - \epsilon ^2\right)}}{6 - \epsilon } & \frac{q^2 \sqrt[3]{36 - \epsilon ^2}}{(\rho \epsilon )^{2/3}} \\
\end{array}
\right) \ , \label{vecteqn5} \\
\end{eqnarray}
\begin{eqnarray}
M_- = \left(
\begin{array}{cc}
 \frac{12 \epsilon  \left(36-\epsilon ^2\right) \rho^2 + q^2 \epsilon ^2 \sqrt[3]{\rho \epsilon  \left(36 - \epsilon ^2\right)} \rho + 2 A q (6 - \epsilon ) \epsilon ^{2/3} \left( 36 - \epsilon ^2\right)^{2/3} \rho^{2/3}}{36 \rho^2 \epsilon } & - \frac{(6 - \epsilon )^2 (\epsilon + 6)}{36 \rho \epsilon } \nonumber\\
  -\frac{12 q^2 \sqrt[3]{\rho \epsilon  \left(36 - \epsilon ^2\right)}}{6 - \epsilon } & \frac{q^2 \sqrt[3]{36 - \epsilon ^2}}{(\rho \epsilon )^{2/3}} \\
\end{array}
\right) \ . \label{vecteqn6} \\
\end{eqnarray}
Here $A = 2 \alpha \left( \Gamma\left( 4/3\right) / \Gamma\left( 1/3 \right)  \right) $, we remind the reader, is essentially the Chern-Simons coupling constant. These mass matrices can now be diagonalized to find the corresponding eigenvalues and are used in the main text.

For completeness, we also include the details of an analogous analysis, for an inherently IR-observer. The metric and the flux are given by
\begin{eqnarray}
&& ds^2 = R^2 \left( -r^2 dt^2 + \frac{dr^2}{r^2} \right) + d\vec{x}^2 \ , \\
&& R^2 = \frac{6}{36 - \epsilon^2} \ , \quad F = \frac{2}{\sqrt{36 - \epsilon^2}} dt \wedge dr \ , \quad {\rm with} \quad \Lambda = - 6 \ .
\end{eqnarray}
Proceeding as above, one obtains the equations of motion for the fluctuations for all sectors. The vector one can now be recast in the following compact form:
\begin{eqnarray}
\nabla_{\rm AdS_2} \Psi_{\pm} = M_{\pm}^2 \Psi_{\pm} \ , \label{vecteqn4}
\end{eqnarray}
where 
\begin{eqnarray}
&& \psi_i = \left( 2 \sqrt{36-\epsilon^2} R^2  \right)  a_i + \partial_r h_{ti} \ , \label{vecteqn1} \\
&& {\rm subsequently} \ , \quad \Psi_{+} = \left\{ a_2 + i a_3 , \psi_2 + i \psi_3 \right\} \ , \label{vecteqn2}  \\
&& {\rm and} \quad  \Psi_{-} = \left\{ a_2 - i a_3 , \psi_2 - i \psi_3 \right\} \ . \label{vecteqn3}
\end{eqnarray}
Also, the mass matrix $M_{\pm}$ is given by
\begin{eqnarray}
&& M_+ = \left( \begin{array}{cc}  12 + \frac{\epsilon^2}{36} \left( q^2 - 12 \right) + 2 \alpha q \sqrt{36 - \epsilon^2 } \left( \frac{\epsilon}{3} - \frac{\epsilon^3}{108} \right)  & - \frac{1}{36} \left( 36 - \epsilon^2 \right)^{3/2} \\ -  \frac{12 q^2}{\sqrt{36 - \epsilon^2}} & q^2  \end{array} \right) \ , \label{vecteqn5} \\
&& M_- = \left( \begin{array}{cc}  12 + \frac{\epsilon^2}{36} \left( q^2 - 12 \right) - 2 \alpha q \sqrt{36 - \epsilon^2 } \left( \frac{\epsilon}{3} - \frac{\epsilon^3}{108} \right)  & - \frac{1}{36} \left( 36 - \epsilon^2 \right)^{3/2} \\ -  \frac{12 q^2}{\sqrt{36 - \epsilon^2}} & q^2  \end{array} \right) \ . \label{vecteqn6}
\end{eqnarray}
%

\section{DBI in Flat space, and Fluctuations}

Let us consider Minkowski background, of the form ${\mathbb R}^{1,1} \times {\mathbb R}^3$--type.
\begin{eqnarray}
ds^2 = - dt^2 + dx^2 + d\vec{y}^2 \ , 
\end{eqnarray}
and consider the following action
\begin{eqnarray}
S = - \tau \int d^5 \xi \sqrt{- {\rm det} \left( \eta + F \right)} + \frac{\alpha\tau}{3!} \int d^5\xi \epsilon^{abcde} A_a F_{bc} F_{de} \ ,
\end{eqnarray}
where $\eta$ is the flat Minkowski metric and $\alpha$ is hitherto free. The equation of motion resulting from the above action is: 
\begin{eqnarray}
&& \partial_a \left( \sqrt{- (\eta + F )} \A^{ab}\right)  + \frac{\alpha}{2} \epsilon^{bcdef} F_{cd} F_{ef} = 0 \ ,  \ , \label{minkgauge} \\
&& {\rm where} \quad \A^{ab} = - \left(\left( \eta + F \right)^{-1} F \left( \eta - F \right)^{-1} \right)^{ab} \ .
\end{eqnarray}
Now, following \cite{Nakamura:2009tf}, we can calculate the equation of motion for a fluctuation field around a classical solution, denoted by $F^{(0)}$. Writing $F = F^{(0)} + f$, we get
\begin{eqnarray}
&& \partial_a \left[ \sqrt{- (\eta + F^{(0)} )} \left( \delta \A^{ab} + \frac{1}{2} {\rm Tr} \left( (\eta + F^{(0)} )^{-1} f \right) F^{(0)ab } \right) \right] + \alpha \epsilon^{bcdef} F_{cd}^{(0)} f_{ef} = 0 \ , \label{fluceqn} \nonumber\\
&& {\rm where} \ , \\
&& \quad \delta \A = - \left( \eta +F \right)^{-1} f \left(\eta - F  \right)^{-1} + \left(\eta + F \right)^{-1} f \left(\eta + F \right)^{-1} F \left( \eta - F \right) ^{-1} \nonumber\\
&& - \left(\eta + F \right)^{-1} F \left(\eta - F \right)^{-1} f \left(\eta - F \right)^{-1} \ .
\end{eqnarray}
The above matrix expression is understood as the tensor with indices raised. The lowered index tensor is obtained by
\begin{eqnarray}
\delta \A_{\rm d} = \eta \left(  \delta \A \right) \eta \ .
\end{eqnarray}
The corresponding equation with lowered indices for the two-form is given by
\begin{eqnarray}
&& \partial^a \left( \O_{ab} \right) +  \alpha \epsilon_b^ {\, \, cdef} F_{cd}^{(0)} f_{ef} = 0 \ , \\
&& \O_{ab} = \sqrt{- (\eta + F^{(0)} )} \left( (\delta \A_{\rm d})_{ab} + \frac{1}{2} {\rm Tr} \left( (\eta + F^{(0)} )^{-1} f \right) F_{ab}^{(0)} \right) \ .
\end{eqnarray}

A simple solution of (\ref{minkgauge}) is given by
\begin{eqnarray}
F^{(0)} = E dt \wedge dx \ , \quad {\rm or} \quad F^{(0)} = H dy^1 \wedge dy^2 \ .
\end{eqnarray}
Here onwards, we will use $\mu, \nu, \ldots$ and $i, j, \ldots$ to represent coordinates along ${\mathbb R}^{1,1}$ and ${\mathbb R}^3$, respectively; on the other hand, $a, b, \ldots$ represent the entire Minkowski background. Taking the electric case, for example, the fluctuation equations divide into two types:
\begin{eqnarray}
&& \partial^a  \O_{a\mu}  = 0 \ , \\
&& \partial^a \O_{a i } + 2\alpha E \epsilon^{01} \epsilon_{i}^{jk} f_{jk} = 0 \  , \label{eqn1} \\
&& {\rm equivalently} \quad \partial^a \O_{a i } - 2\alpha E  \epsilon_i^{jk} f_{jk} = 0 \ . \label{eqn2}
\end{eqnarray}
We are further taking $\epsilon_{01} = 1$ and $\epsilon_{234}=1$. We also have the Bianchi identity:
\begin{eqnarray}
\partial_a f_{bc} + \partial_b f_{ca} + \partial_c f_{ab} = 0 \ .
\end{eqnarray}
In order to write the equations in terms of $f_{ij}$, we note that
\begin{eqnarray}
&& \O_{\mu i} = - \frac{1}{\sqrt{1- E^2}} f_{\mu i} \ ,  \quad \O_{ij} = - \sqrt{1 - E^2} f_{ij } \ , \\
&& \O_{\mu\nu} = - \frac{1}{\left( 1 - E^2 \right)^{3/2}} f_{\mu\nu} \ .
\end{eqnarray}

The equation of motion for fluctuations in (\ref{eqn2}) can now be written as:
\begin{eqnarray}
\left[ \partial_\mu \partial^\mu + \left( 1 - E^2 \right) \partial_j \partial^j \right] \tilde{f}_i + \left( 4 \alpha E \sqrt{1 - E^2}\right) \epsilon_i^{\, \, jk} \partial_j \tilde{f}_k = 0 \ , \label{redeqndbi}
\end{eqnarray}
where we have used the following:
\begin{eqnarray}
&& \tilde{f}_i = \frac{1}{2}\epsilon_i^{\, \, jk} f_{jk} \quad \implies \quad f_{ij} = \epsilon_{ij}^{\, \, \, \, k} \tilde{f}_k \ . \\
&& {\rm Also} \quad \partial^j \tilde{f}_j = \frac{1}{2} \epsilon_j^{\, \, kl} \partial^j f_{kl} = 0 \ .
\end{eqnarray}
The last line above is obtained by using the anti-symmetry property of the epsilon symbol and the Bianchi identity.

To determine the corresponding mass spectrum, in close resemblance to what is discussed in the previous appendix, let us write 
\begin{eqnarray}
&& \partial_\mu \partial^\mu \tilde{f}_i =  - p^2 \tilde{f}_i \ , \quad  \partial_j \partial^j \tilde{f}_i = - k^2 \tilde{f}_ i \ , \\
&& {\rm with} \quad \tilde{f}_i = c_i e^{i p_\mu x^\mu + i k_j x^j} \ .
\end{eqnarray}
The equations of motion in (\ref{redeqndbi}) now yields the following algebraic conditions:
\begin{eqnarray}
&& - \Delta^2 \tilde{f}_1 + i \left( 4 \alpha E \sqrt{1 - E^2}  \right) \left[ k_2 \tilde{f}_3 - k_3 \tilde{f}_2 \right] = 0 \ , \\
&&  - \Delta^2 \tilde{f}_2 + i \left( 4 \alpha E \sqrt{1 - E^2} \right) \left[ k_3 \tilde{f}_1 - k_1 \tilde{f}_3 \right] = 0 \ , \\
&& - \Delta^2 \tilde{f}_3 + i \left( 4 \alpha E \sqrt{1 - E^2} \right) \left[ k_1 \tilde{f}_2 - k_2 \tilde{f}_1 \right] = 0 \ , \\
&&  {\rm with} \quad \Delta^2 =  p^2 + \left( 1 - E^2 \right) k^2 \ .
\end{eqnarray}
This yields
\begin{eqnarray}
&& \Delta^2 = \pm \left( 4 \alpha E \sqrt{1- E^2} \right) k \ , \\
&& \implies  \quad m^2 = - p_\mu p^\mu = \left( 1- E^2 \right) \left[ k^2 - \frac{4 \alpha E k}{\sqrt{1 - E^2}} \right] \ . 
\end{eqnarray}
Thus a tachyonic mode is obtained in the range $0< k < \left(4 \left| \alpha E\right| \right) / \sqrt{1 - E^2 } $. It is now clear that, by tuning $E \to 1$, the entire momentum-mode can be made unstable. Thus, even though there is an upper limit of the electric field, the entire momentum-space is potentially unstable.

\end{document}